\documentclass[conference,onecolumn]{IEEEtran}

\usepackage{cite}      

\usepackage{graphicx}  

\usepackage{indentfirst}
\usepackage{amssymb}

\begin{document}

\title{Opportunistic Relay Selection with Limited Feedback}

\author{Caleb K. Lo, Robert W. Heath, Jr. and Sriram Vishwanath \\Wireless Networking and Communications Group \\ Department of Electrical and Computer Engineering \\ The University of Texas at Austin, Austin, Texas 78712-0240 \\ Email: [clo, rheath, sriram]@ece.utexas.edu
\thanks{Caleb K. Lo was supported by a Microelectronics and Computer Development (MCD) Fellowship and a Thrust 2000 Endowed Graduate Fellowship through The University of Texas at Austin.  Robert Heath was supported in part by the National Science Foundation under grant CNS-626797, the Office of Naval Research under grant number N00014-05-1-0169, and the DARPA IT-MANET program, Grant W911NF-07-1-0028.   Sriram Vishwanath was supported by the National Science Foundation under grants CCF-055274, CCF-0448181, CNS-0615061 and CNS-0626903.}}
\date{}


%


\renewcommand{\baselinestretch}{2}

\twocolumn
\renewcommand{\baselinestretch}{1}

\setcounter{page}{1}
\newpage

\maketitle

\begin{abstract}
It has been shown that a decentralized relay selection protocol based on opportunistic feedback from the relays yields good throughput performance in dense wireless networks.  This selection strategy supports a hybrid-ARQ transmission approach where relays forward parity information to the destination in the event of a decoding error.  Such an approach, however, suffers a loss compared to centralized strategies that select relays with the best channel gain to the destination.  This paper closes the performance gap by adding another level of channel feedback to the decentralized relay selection problem.  It is demonstrated that only one additional bit of feedback is necessary for good throughput performance.  The performance impact of varying key parameters such as the number of relays and the channel feedback threshold is discussed.  An accompanying bit error rate analysis demonstrates the importance of relay selection.
\end{abstract}


%

\section{Introduction}
Message forwarding in multihop networks occurs over inherently lossy wireless links and coding strategies are needed to meet the network QoS requirements.  Hybrid-ARQ is such a coding strategy that is especially effective in dense networks, as intermediate nodes can act as relays, forwarding parity information to the destination.  If the destination detects uncorrectable packet errors and broadcasts a retransmission request to the network, the relays are well-positioned to transmit parity information more reliably than the source.  

Relay selection techniques have been studied extensively in recent years \cite{CheSerETAL:DistPoweAlloPara:Nov:05,LinErk:RelaSearAlgoCode:Nov:05,SadHanETAL:DistRelaAssiAlgo:Jun:06,ZhaAdvETAL:ImprAmplForwRela:Jul:06,SreYanETAL:RelaSeleStraCell:Oct:03,BleKhiETAL:SimpCoopDiveMeth:Mar:06,LuoBluETAL:ApprCoopMultAnte:Sep:04,ZhaVal:PracRelaNetwGene:Jan:05}.  In our previous work on this topic, we proposed a decentralized selection strategy for relay selection in dense mesh networks \cite{LoETAL:HybrARQMultNetw:Apr:07}, where decoding relays contend to forward parity information to the destination using rate-compatible punctured convolutional (RCPC) codes \cite{Hag:RateCompPuncConv:Apr:88}.

Our random access-based approach, which is based on opportunistic feedback \cite{TanHea:OppoFeedDownMult:Oct:05}, is distinct from centralized strategies that select the relay with the best instantaneous channel gain to the destination \cite{LuoBluETAL:ApprCoopMultAnte:Sep:04,ZhaVal:PracRelaNetwGene:Jan:05}.  Such centralized strategies, though, yield better throughput than our decentralized approach.  

To close this performance gap, we propose a refinement of our selection strategy via channel feedback.  In our previously proposed approach, if a decoding relay successfully sends a ``Hello'' message to the source in a minislot, it is declared to be the ``winner'' for that minislot.  The source then randomly chooses a relay among the set of all ``winners.''  In this paper, we refine the relay selection among the set of all ``winners'' by biasing the selection towards those relays with channel gains to the destination that are above a particular threshold.  For example, if the set of ``winners'' consists of one relay with a channel gain above the threshold and one relay with a channel gain below the threshold, the relay with a channel gain above the threshold is more likely to be chosen by the source than the other relay.

We briefly discuss how our previously proposed relay selection strategy differs from the notion of multiuser diversity \cite{QinBer:DistApprExplMult:Feb:06, VisETAL:OppoBeamDumbAnte:Jun:02}.  The basic premise behind multiuser diversity is that in a system with many users with independently fading channels, the probability that at least one user will have a ``good'' channel gain to the transmitter is high.  Then, the user with the best channel gain to the transmitter can be serviced, which will yield the maximum throughput.  In our setup, the analogous approach would be to always choose the decoding relay that has the best channel gain to the destination to forward parity information.  Our decentralized approach, though, allows any decoding relay to have a chance of being selected to forward parity information as long as it sends at least one ``Hello'' message to the source and wins at least one minislot.

\section{System Model}
\begin{figure}[t]
\begin{center}
\includegraphics[width=3.0in]{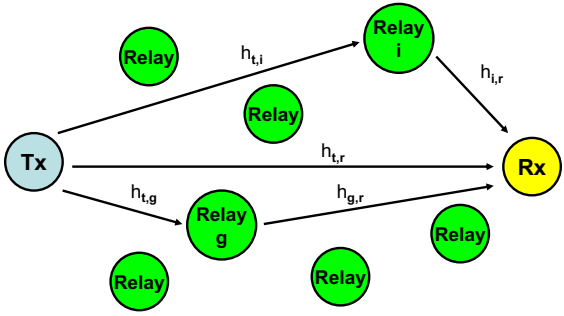}
\end{center}
\caption{Relay network.}
\label{system-model}
\end{figure}

Consider the setup in Fig. \ref{system-model}.  Each relay operates in a half-duplex mode and is equipped with a single antenna.  We use boldface notation for vectors.  SNR represents the signal-to-noise ratio.  $|h|^2$ denotes the absolute square of $h$.  $Q(\cdot)$ is the standard Q-function, and $Pr(X \leq x)$ denotes the probability that a realization of the random variable $X$ is at most $x$.

Transmission occurs over a set of time slots $\{t_1,...,t_m\}$ which are of equal duration.  We use the ARQ/FEC strategy in \cite{Hag:RateCompPuncConv:Apr:88}.  Initially, the source has a k-bit message $\textbf{w}$ that is encoded as an n-bit codeword $\textbf{x}$.  The source chooses code-rates from a RCPC code family, say $\{R_1,R_2,...,R_m\}$ where $R_1 > R_2 > \cdots > R_m$.

Before $t_1$, the source and destination perform RTS/CTS-based handshaking to achieve synchronization.  During $t_1$, the source transmits a subset $\textbf{x}_1$ of the bits in $\textbf{x}$ such that $\textbf{x}_1$ forms a codeword from the rate-$R_1$ code.  The destination observes
\begin{equation}
\textbf{y}_{r,1} = h_{t,r}\textbf{x}_1 + \textbf{n}_{r,1}
\end{equation}
while relay $i \in \{1,2,...,K_r\}$ observes
\begin{equation}
\textbf{y}_{i,1} = h_{t,i}\textbf{x}_1 + \textbf{n}_{i,1}.
\end{equation}
Here, $h_{t,i}$ is a Rayleigh fading coefficient for the channel between the source and node $i$, while $\textbf{n}_{i,j} $ is additive white Gaussian noise with variance $N_0$ at node $i$ during time slot $t_j$.  We assume that all fading coefficients are constant over a time slot and vary from slot to slot; we also assume that fading and additive noise are independent across all nodes.  In addition, we assume that all nodes have no prior knowledge of fading coefficients and use training data to learn them.

The destination attempts to decode $\textbf{y}_{r,1}$.  If decoding is successful, the destination broadcasts an ACK message to all of the relays and the source.  Otherwise, the destination broadcasts a NACK message; the source now has to select one of the relays to forward additional parity information that will assist the destination in recovering $\textbf{w}$.

\section{Relay Selection Via Limited Feedback}\label{rel-sel}
\begin{figure}[t]
\begin{center}
\includegraphics[width=3.0in]{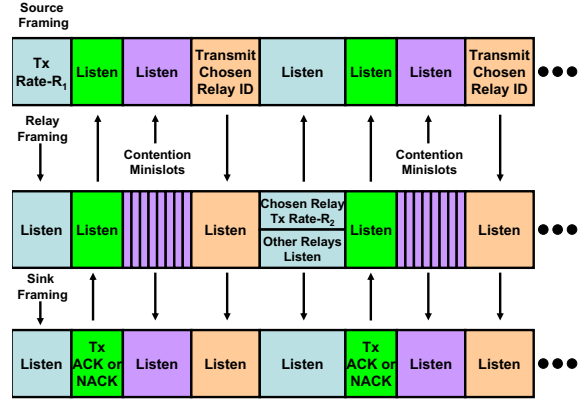}
\end{center}
\caption{Framing structure for decentralized selection strategy.}
\label{framing-structure}
\end{figure}

We briefly review our proposed relay selection strategy in \cite{LoETAL:HybrARQMultNetw:Apr:07}.  The framing structure for our relay selection strategy is shown in Fig. \ref{framing-structure}.  We assume in Fig. \ref{framing-structure} that the destination sends a NACK after $t_1$ and $t_2$ to trigger the relay contention process.  Let $\mathcal{R}_{sel}$ denote the set of relays that can participate in the contention process.  If relay $i \in \mathcal{R}_{sel}$, then relay $i$ has successfully recovered $\textbf{w}$ and has a channel gain to the destination $|h_{i,r}|^2$ that is above a threshold $\eta_{opp}$.  Relay $i$ can determine $|h_{i,r}|^2$ by listening to the destination's NACK, which is embedded in a packet that contains training data.

All relays in $\mathcal{R}_{sel}$ use the same $K$ minislots for feedback to the source.  During minislot $b$, each relay $i \in \mathcal{R}_{sel}$ sends a ``Hello'' message to the source with probability $p_i$.  We refer to this approach as a $\textit{1-bit}$ strategy, where the ``Hello'' message is an ID number that has been assigned to each relay.  Successful contention occurs during minislot $b$ if exactly one relay $i \in \mathcal{R}_{sel}$ sends a ``Hello'' message.  The source then declares that relay as the ``winner'' for minislot $b$.  After minislot $K$, the source randomly selects one of the ``winners'' $i_{t}$; if there are no ``winners,'' the source will transmit during $t_2$.

In this work, we modify the $\textit{1-bit}$ strategy by appending a check bit to the ``Hello'' message; the check bit is set to `1' only if $|h_{i,r}|^2 > \beta_{opp}$ for $\beta_{opp} > \eta_{opp}$.  Again, successful contention occurs during minislot $b$ if exactly one relay $i \in \mathcal{R}_{sel}$ sends a ``Hello'' message.  We refer to this approach as a $\textit{2-bit}$ strategy.   

After minislot $K$, if either all of the ``winners'' sent a check bit of '0', all of the winners sent a check bit of '1', or there are no ``winners,'' the $\textit{2-bit}$ strategy reduces to the $\textit{1-bit}$ strategy.  Otherwise, the source will randomly select one of the ``winners'' $i_{t}$ that sent a check bit of '1' with probability $q > 0.5$; one of the ``winners'' $i_{t}$ that sent a check bit of '0' is randomly selected with probability $1-q$.  Thus, the $\textit{2-bit}$ strategy refines the $\textit{1-bit}$ strategy by further biasing the selection process in favor of the relays with the best channel gains to the destination. 

During $t_2$, relay $i_{t}$ (or the source) transmits a subset $\textbf{x}_2$ of the bits in $\textbf{x}$ such that $\textbf{x}_1 \cup \textbf{x}_2$ forms a codeword from the rate-$R_2$ code in the RCPC family.  The destination combines $\textbf{y}_{r,1}$ with
\begin{equation}
\textbf{y}_{r,2} = h_{i_{t},r}\textbf{x}_2 + \textbf{n}_{r,2}
\end{equation}
and attempts to decode $\textbf{y}_{r,1} \cup \textbf{y}_{r,2}$ based on the rate-$R_2$ code.  If unsuccessful decoding occurs again, the destination broadcasts another NACK and the contention process repeats until either the destination successfully recovers $\textbf{w}$ or the rate-$R_m$ code has been used without successful decoding.

To compute the dimensionless effective rate $R_{avg}$ of this strategy, we use \cite[equation (16)]{Hag:RateCompPuncConv:Apr:88}
\begin{equation}\label{hagenauer-throughput}
R_{avg} = \frac{k}{n+M}\cdot \frac{P}{P+l_{AV}}
\end{equation}
where $l_{AV}$ is the average number of additionally transmitted bits per $P$ information bits.  Here, $M$ is the memory of the mother code for the RCPC family.  We refer to $R_{avg}$ as the throughput of this strategy in the rest of this paper. 

For simulation purposes, we employ the path loss model described in \cite{ZhaVal:PracRelaNetwGene:Jan:05}; thus, the received energy at node $i$ is
\begin{eqnarray}
\mathcal{E}_i & = & |h_{b,i}|^2\mathcal{E}_{x} \\
& = & (\lambda_c/4\pi d_0)^2(d_{b,i}/d_0)^{-\mu}\mathcal{E}_{x}
\end{eqnarray}
where $\mathcal{E}_{x}$ is the energy in the transmitted signal $\textbf{x}$.  Here, $\lambda_c$ is the carrier wavelength, $d_0$ is a reference distance, $d_{b,i}$ is the distance between transmitting node $b$ and receiving node $i$, and $\mu$ is a path loss exponent.

We adopt similar simulation parameters as those in \cite{ZhaVal:PracRelaNetwGene:Jan:05}.  Here, we employ a carrier frequency $f_c$ = 2.4GHz, $d_0$ = 1m, $d_{t,r}$ = 100m and $\mu$ = 3, where $d_{t,r}$ is the distance between the source and the destination.  We then uniformly distribute $K_r = 20$ relays in the region between the source and the destination such that each relay $i$ is $d_{i,r} < d_{t,r}$ units from the destination.  We also use the WiMAX signaling bandwidth, which is roughly 9 MHz \cite{WireMANWorkGrp}; given a noise floor of -204 dB/Hz this yields a noise value $N_0 = -134$ dB.  BPSK modulation is used for all packet transmissions, and all of the relays and the destination use ML decoding.

We use the codes of rates $\{4/5, 2/3, 4/7, 1/2, 1/3\}$ from the $M = 6$ RCPC family in \cite{Hag:RateCompPuncConv:Apr:88}.  We perform concatenated coding, where the outer code is a (255, 239) Reed-Solomon code with symbols from $GF(2^8)$; this code can correct at most 8 errors.  The mother code for the RCPC family is a rate-1/3 convolutional code with constraint length 7 and generator polynomial (145 171 133) in octal notation.

In this section and in Section \ref{perf-impact}, we define the average received SNR at the destination as follows.  Assume that the source uses a transmit energy of $\mathcal{E}_{t}(\gamma)$ during time slot $t_1$ that yields an average SNR $\gamma$ at the destination; then, all transmitting nodes will use a transmit energy of $\mathcal{E}_{t}(\gamma)$ during all subsequent transmission cycles.

\begin{figure}[tb]
\begin{center}
\includegraphics[width=3.0in]{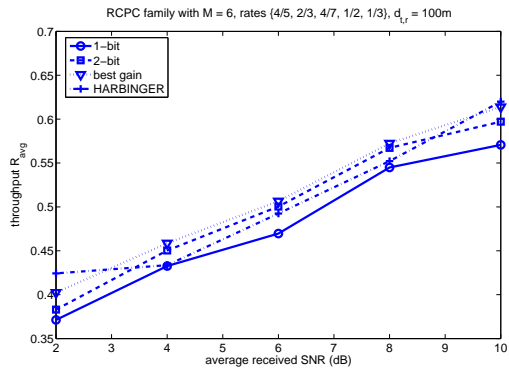}
\end{center}
\caption{Comparison of 1-bit and 2-bit feedback strategies.}
\label{throughput2}
\end{figure}

Fig. \ref{throughput2} compares the throughput yielded by the $\textit{1-bit}$ and $\textit{2-bit}$ strategies.  We also plot the throughput yielded by the GPS-based HARBINGER method \cite{ZhaVal:PracRelaNetwGene:Jan:05} and by a centralized strategy that always selects the decoding relay with the best instantaneous channel gain to the destination to forward parity information.  We have $K = 10$ minislots.  For the $\textit{1-bit}$ and $\textit{2-bit}$ strategies, we set $\eta_{opp} = -91 dB$; we also set $\beta_{opp} = -86 dB$.  We set the feedback probability $p_i = 0.3$ for both strategies.  In addition, we set the ``winner'' selection probability $q = 0.75$ for the $\textit{2-bit}$ strategy.  We see that the $\textit{2-bit}$ strategy closes the performance gap between the $\textit{1-bit}$ strategy and the centralized approach.  Thus, using a limited amount of channel feedback improves the performance of our relay selection strategy.

We also observe that the $\textit{1-bit}$ and $\textit{2-bit}$ strategies offer comparable performance to the HARBINGER method.  Note that the $\textit{2-bit}$ strategy outperforms the HARBINGER method for some values of the received SNR.  The intuition behind this result is that the HARBINGER method optimizes the average received SNR at the destination by selecting the closest decoding relay to the destination.  This method, though, does not necessarily select the ``best'' decoding relay that has the highest instantaneous channel gain to the destination.  Also, the inherent randomness of the $\textit{1-bit}$ and $\textit{2-bit}$ strategies allows for the possibility of choosing the ``best'' decoding relay.  Thus, the HARBINGER method does not necessarily outperform our selection strategies for all received SNR values.

\section{Performance Impact of Varying System Parameters}\label{perf-impact}
A joint optimization of all of the key system parameters would enable computation of the maximum throughput yielded by the $\textit{1-bit}$ and $\textit{2-bit}$ strategies.  This optimization, though, is fairly difficult to perform; instead, in this section we provide some insight for system designers by varying some of the key parameters in isolation and illustrating the resulting impact on the throughput.

\begin{figure}[tb]
\begin{center}
\includegraphics[width=3.0in]{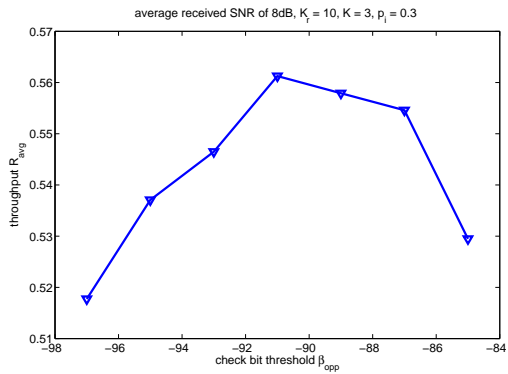}
\end{center}
\caption{Throughput as a function of check bit threshold.}
\label{beta-thresh}
\end{figure}

Fig. \ref{beta-thresh} illustrates the throughput of the $\textit{2-bit}$ strategy for various values of the check bit threshold $\beta_{opp}$.  Here we have $K_r = 10$ relays and $K = 3$ minislots.  We have an average received SNR at the destination of 8dB.  We see that if $\beta_{opp}$ is close to $\eta_{opp}$, the performance of the $\textit{2-bit}$ strategy suffers since the $\textit{2-bit}$ strategy essentially reduces to the $\textit{1-bit}$ strategy.  Also, we see that if $\beta_{opp}$ is too large, the performance of the $\textit{2-bit}$ strategy suffers.  This is because the probability of selecting a decoding relay $i$ such that $|h_{i,r}|^2 > \beta_{opp}$ decreases as $\beta_{opp}$ increases, which causes the $\textit{2-bit}$ strategy to reduce to the $\textit{1-bit}$ strategy again.  Thus, it is apparent that there is an optimal value of $\beta_{opp}$ for each value of the average received SNR that maximizes the throughput of the $\textit{2-bit}$ strategy.

\begin{figure}[b]
\begin{center}
\includegraphics[width=3.0in]{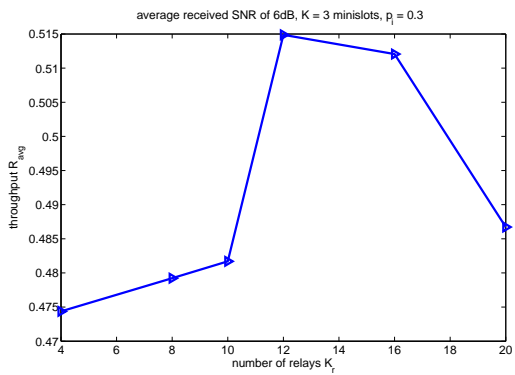}
\end{center}
\caption{Throughput as a function of number of relay nodes.}
\label{relay-number}
\end{figure}

Fig. \ref{relay-number} illustrates the throughput of the $\textit{1-bit}$ strategy for a varying number of relay nodes.  We have $K = 3$ minislots and an average received SNR of 6dB at the destination.  We see that there is an optimal number of relay nodes for which the throughput is maximized.  Note that if the number of relay nodes is small, the probability that any of them decode the source message and send a ``Hello'' message to the source is also small.  On the other hand, if the number of relay nodes is large, the probability that at least two relays decode the source message and attempt to send a ``Hello'' message to the source in each minislot is also large; thus, a collision is likely to occur in each minislot.

\begin{figure}[tb]
\begin{center}
\includegraphics[width=3.0in]{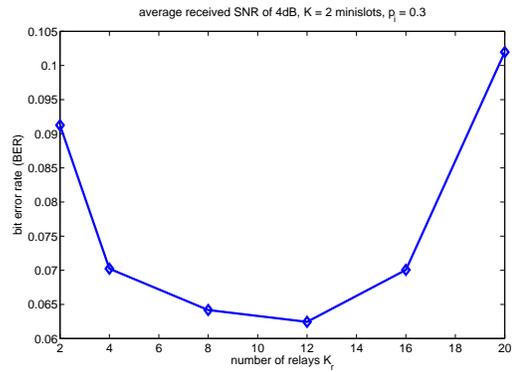}
\end{center}
\caption{Bit error rate as a function of number of relay nodes.}
\label{ber-relay-number}
\end{figure}

Fig. \ref{ber-relay-number} also illustrates the effect on the performance of the $\textit{1-bit}$ strategy of varying the number of relay nodes.  Instead of considering the throughput, though, we consider the bit error rate (BER); we focus on transmission during time slot $t_2$ where the rate-2/3 code from the RCPC family is used.  Here we have $K = 2$ minislots and we set the feedback threshold $\eta_{opp} = -98 dB$.  Again, we notice that the performance of the $\textit{1-bit}$ strategy suffers when the number of relay nodes is either small or large.

\section{BER Analysis}
Assume that we employ Viterbi decoding at the relays and at the destination.  Recall that $P$ is the puncturing period of the RCPC family.  Let $P_d$ be the probability that an incorrect path of weight $d$ is selected by the Viterbi decoder, and let $d_{free}$ be the free distance of the member of the RCPC family that is currently being used for decoding.  Also, let $c_d$ be the total number of non-zero information bits on all paths of weight $d$.  From \cite[equation (9)]{Hag:RateCompPuncConv:Apr:88} we see that the bit error rate $P_b$ can be upper-bounded as
\begin{equation}\label{ber-ub}
P_b \leq \frac{1}{P}\sum_{d=d_{free}}^{\infty}c_dP_d.
\end{equation}
Let $\gamma_r$ denote the received SNR at the destination.  Since we are essentially dealing with a binary-input AWGN channel with binary output quantization, we use \cite[equation (12.39b)]{LinCos:ErroContCodi:04} to see that $P_b$ can be further upper-bounded as
\begin{eqnarray}
P_b & < & \frac{1}{P}\sum_{d=d_{free}}^{\infty}c_d\Big(2\sqrt{p(1-p)}\Big)^d \\
& = & \frac{1}{P}\sum_{d=d_{free}}^{\infty}c_d\cdot \label{ber-upper-bound} \\
& & \bigg(2\sqrt{Q\Big(\sqrt{2\gamma_r}\Big)\Big(1-Q\Big(\sqrt{2\gamma_r}\Big)\Big)}\bigg)^d. \nonumber
\end{eqnarray}

Since $g(\gamma_r) = Q(\sqrt{2\gamma_r})(1-Q(\sqrt{2\gamma_r}))$ is a monotonically decreasing function for non-negative $\gamma_r$, we see that $P_b$ monotonically decreases for increasing values of the received SNR.  This demonstrates the utility of relay selection, as transmission from relay nodes will yield a higher average received SNR at the destination than transmission from the source.

To illustrate this point, consider the following simple example.  We have the same simulation parameters as in Section \ref{rel-sel}, except that now we have $K_r = 1$ relay, $K = 1$ minislot and a feedback probability $p_i = 1$.  We place this relay at a location that is 25 meters from the source and 75 meters from the destination.  During time slot $t_1$, the source uses a transmit energy that is 101dB above the noise floor $N_0$, which yields an average received SNR at the destination of $\gamma_{t,r} = 0.952dB$.  We consider the $\textit{1-bit}$ strategy here and set $\eta_{opp} = -91dB$.

Consider time slot $t_2$, where we assume that the destination did not successfully recover $\textbf{w}$ during $t_1$.  Now, if the relay is selected to forward parity information during $t_2$, the average received energy at the destination is
\begin{eqnarray}
\mathcal{E}_r & = & \bigg(\frac{3\cdot10^8}{2.4\cdot10^9}\cdot\frac{1}{4\pi}\bigg)^2\bigg(\frac{1}{75}\bigg)^{-3}10^{(-134+101)/10} \nonumber \\
& \approx & 1.17\cdot10^{-13}. \nonumber
\end{eqnarray}
Thus, we have an average received SNR at the destination of $\gamma_{1,r} = \mathcal{E}_r/N_0 \approx 4.7dB.$

From \cite{Hag:RateCompPuncConv:Apr:88} we can determine the bit weight enumerating function (WEF) weights $c_d$ for the rate-2/3 code from the RCPC family.  In particular, we see from \cite[Table II(c)]{Hag:RateCompPuncConv:Apr:88} that the only non-zero values of $c_d$ are \[c_d = \{12,280,1140,5104,24640,108512\}\] for $d = \{6,7,8,9,10,11\}$.  Now we substitute these values of $c_d$ and $d$ along with $\gamma_r = \gamma_{1,r}$ into (\ref{ber-upper-bound}).  We find that the BER $P_b$ is upper-bounded as $P_b < 5.42\cdot10^{-4}$.

Since $Pr(\gamma_r < \gamma_{1,r}) = 0.368$, we want to evaluate the performance of our selection strategy for a wider range of $\gamma_r$.  In particular, we find that $Pr(\gamma_r < 2) = 0.492$; if we substitute $\gamma_r = 2$ into (\ref{ber-upper-bound}) we have $P_b < 0.0688$.

On the other hand, assume that the source forwards parity information during $t_2$.  If we substitute $\gamma_r = \gamma_{t,r}$ into (\ref{ber-upper-bound}), we find that the BER $P_b$ is upper-bounded as $P_b < 5.55$.  

Again, since $Pr(\gamma_r < \gamma_{t,r}) = 0.368$, we evaluate the performance of this approach for a wider range of $\gamma_r$.  We find that $Pr(\gamma_r < 0.85) = 0.495$; if we substitute $\gamma_r = 0.85$ into (\ref{ber-upper-bound}) we have $P_b < 64.7$.  Thus, it is apparent that relay selection leads to significant gains in BER performance.

Since relaying leads to significantly improved BER performance, we want to determine the probability of relay selection for this example.  Here, the relay is selected if it recovers $\textbf{w}$ in $t_1$ and has a channel gain to the destination $|h_{1,r}|^2 > \eta_{opp}$.

Recall our assumption that all channels in our setup undergo Rayleigh fading.  First, the probability that the relay has a sufficiently high channel gain to the destination is 
\begin{eqnarray}
P_2 & = & \int_{\eta_{opp}}^{\infty}\frac{1}{\gamma_{1,r}}e^{-\chi/\gamma_{1,r}}d\chi \\
& \approx & 1. \nonumber
\end{eqnarray}

Thus, we only have to consider the probability $P_1$ that the relay recovers $\textbf{w}$ in time slot $t_1$.  From \cite[equation (20)]{Hag:RateCompPuncConv:Apr:88}, the probability $P_{\textnormal{err}}$ that the relay cannot recover $\textbf{w}$ in $t_1$ is
\begin{equation}
P_{\textnormal{err}} < 1 - \bigg(1-\frac{1}{P}\sum_{d=d_{free}}^{\infty}c_dP_d\bigg)^{n+M}
\end{equation}
where the non-zero values of $c_d$ are for the rate-4/5 code from the RCPC family.  By using (\ref{ber-upper-bound}), $P_1$ is lower-bounded as 
\begin{eqnarray}
P_1 & = & 1 - P_{\textnormal{err}} \nonumber \\
& = & (1-P_b)^{n+M} \\
& > & \bigg(1-\frac{1}{P}\sum_{d=d_{free}}^{\infty}c_d\cdot \label{p1-lower-bound} \\
& & \bigg(2\sqrt{Q\Big(\sqrt{2\gamma_r}\Big)\Big(1-Q\Big(\sqrt{2\gamma_r}\Big)\Big)}\bigg)^d\bigg)^{n+M} \nonumber 
\end{eqnarray}
In particular, we see from \cite[Table II(c)]{Hag:RateCompPuncConv:Apr:88} that the only non-zero values of $c_d$ are \[c_d = \{24,376,3464,30512,242734,1890790\}\] for $d = \{4,5,6,7,8,9\}$.  We have $P = 8$, $n = 2040$, $M = 6$ and the average received SNR at the relay during time slot $t_1$ is $\gamma_{t,1} \approx 19dB$.  If we substitute these values of $P$, $n$, $M$ and $\gamma_r = \gamma_{t,1}$ into (\ref{p1-lower-bound}) we see that $P_1 \approx 1$.  

Again, since $Pr(\gamma_r < \gamma_{t,1}) = 0.368$, we evaluate the performance of our selection strategy for a wider range of $\gamma_r$.  In particular, we find that $Pr(\gamma_r < 5) = 0.0608$; if we substitute $\gamma_r = 5$ into (\ref{p1-lower-bound}) we see that $P_1 > 0.851$.  Thus, we have a good chance of reaping the benefits of relay selection.

\section{Conclusion}
In this paper we presented a strategy for improving the throughput of our previously proposed decentralized relay selection protocol.  We modified our protocol by using a limited amount of channel feedback to close the performance gap between our protocol and centralized strategies that select the relay with the best channel gain to the destination.  To understand the performance impact of different system parameters, we presented simulation results and discussed their applicability to system design.  We performed a simple BER analysis to further illustrate the gains achieved by relaying.
  




%

\end{document}